# Spatial profiles of photon chemical potential in near-field thermophotovoltaic cells

Dudong Feng[1], Eric J. Tervo[2,a)], Dragica Vasileska[3], Shannon K. Yee[1], Ajeet Rohatgi[4], and Zhuomin M. Zhang[1,a)]

[1] George W. Woodruff School of Mechanical Engineering, Georgia Institute of Technology, Atlanta, GA 30332

[2] National Renewable Energy Laboratory, 15013 Denver West Parkway, Golden, CO 80401

[3] School of Electrical, Computer and Energy Engineering, Arizona State University, Tempe, AZ 85287

[4] School of Electrical and Computer Engineering, Georgia Institute of Technology, Atlanta, GA 30332

[a)] Authors to whom correspondence should be addressed:
zhuomin.zhang@me.gatech.edu; Eric.Tervo@nrel.gov

**Abstract**

Emitted photons stemming from the radiative recombination of electron-hole pairs carry chemical potential in radiative energy converters. This luminescent effect can substantially alter the local net photogeneration in near-field thermophotovoltaic cells. Several assumptions involving the luminescent effect are commonly made in modeling photovoltaic devices; in particular, the photon chemical potential is assumed to be zero or a constant prescribed by the bias voltage. The significance of photon chemical potential depends upon the emitter temperature, the semiconductor properties, and the injection level. Hence, these assumptions are questionable in thermophotovoltaic devices operating in the near-field regime. In the present work, an iterative solver that combines fluctuational electrodynamics with the drift-diffusion model is developed to tackle the coupled photon and charge transport problem, enabling the determination of the spatial profile of photon chemical potential beyond the detailed balance approach. The difference between the results obtained by allowing the photon chemical potential to vary spatially and by assuming a constant value demonstrates the limitations of the conventional approaches. This study is critically important for performance evaluation of near-field thermophotovoltaic systems.

## I. INTRODUCTION

Thermal radiation is often characterized as a photon gas with zero chemical potential because the number of photons is not conserved in a blackbody enclosure. However, in general, photons obey the Bose-Einstein statistics with a chemical potential and can interact with other quasiparticles [1-3]. For a semiconductor, the emission or absorption of photons whose energies exceed the bandgap energy ($E_g$) are associated with the radiative recombination or photogeneration processes, in which an electron-hole pair is either eliminated or created. From a thermodynamics point of view, the Gibbs free energy should be conserved in these processes. Thus, the photon chemical potential ($\mu$) should equal the difference between those of the electrons and holes [4]. The emitted photon flux may be higher or lower than that from a blackbody at the same temperature, depending on whether $\mu$ is positive or negative, resulting in luminescence or negative luminescence effects [3]. The luminescence effect is essential to the operation of light-emitting diodes (LEDs) [2]. Furthermore, photovoltaics (PVs), thermophotovoltaics (TPVs), and thermoradiative cells (TRs) may also be subject to nontrivial electroluminescence due to the radiative recombination of injected electrons and holes [3-5].

As shown in Fig. 1, in an ideal scenario, an incident photon above the bandgap energy is absorbed by a TPV cell, which generates one electron-hole pair. Some of these free carriers are separated in the depletion region and flow in opposite directions, providing electrical power to an external load. Meanwhile, some of these excess electrons and holes recombine and generate luminescent photons, which carry a chemical potential equal to the difference between the quasi-Fermi levels of the electrons and the holes: $E_{f,e}$ and $E_{f,h}$. For far-field TPVs operating at high emitter temperatures (> 1500 K), the flux of luminescent photons is usually negligible compared to the high incident photon flux. Nevertheless, when the separation distance between the emitter and the cell in a TPV system reaches the subwavelength regime, the radiative exchange is greatly enhanced due to photon tunneling. The effect of photon chemical potential can be significant for near-field TPV cells when the enhanced luminescent emission intensity becomes comparable to the incident intensity originated from a moderate-temperature emitter [4, 6]. Therefore, careful consideration of the photon chemical potential profile is crucial for prediction of the performance of near-field TPV systems.

Near-field TPV systems hold promise for energy harvesting due to the greatly enhanced power throughput with a potential improvement in conversion efficiency [7]. Significant theoretical and experimental advances have been made in recent years [8-11]. Most early studies on TPV systems did not consider the effect of luminescence from the TPV cell [12-14], or used simplified models for the radiative recombination [15] as well as reabsorption (also called photon recycling) [16]. The consideration of radiative recombination and reabsorption is an approximate approach of treating luminescence and photon recycling effects without explicitly using photon chemical potential in the Bose-Einstein statistics [17, 18]. External luminescence also affects the net radiative transfer rate between the emitter and the cell, especially when the emitter is at moderate temperatures [19-22]. These studies used a direct modeling method for photon exchange between the emitter and the cell by assuming a spatially uniform photon chemical potential that is equal to the elementary charge ($q$) times the operating voltage ($V$); furthermore, the detailed balance approach was applied to calculate the carrier concentrations and thus the recombination rates. Feng *et al*. [23] illustrated the near-field effect on the dark current due to the modified saturation current in the near-field regime under the same assumptions. When the photogeneration rate is very high, the injected concentration could be comparable to or higher than the doping concentration. As shown by Blandre *et al*. [24] based on the full drift-diffusion model, the quasi-Fermi levels split nonuniformly within the active region of the cell. However, their study used a radiative recombination model to treat the luminescence effect without considering photon recycling and the effect of photon chemical potential on the net radiative heat transfer.

This work describes an iterative method that combines fluctuational electrodynamics with the full drift-diffusion model to solve the coupled charge and photon transport equations. The focus is on the determination of the impact of the photon chemical potential on the performance of the TPV systems. By comparison of the solution using the iterative method with that using the detailed balance approach based on a constant (or zero) photon chemical potential, the error caused by these assumptions can be quantified for different TPV systems. Two InAs near-field TPV cells with different thicknesses are selected as examples to illustrate the profile of photon chemical potential and when the assumption of a constant value may break down. The effects of injection level and surface recombination on the width and band structure of the depletion region are examined. For both the thin and thick TPV systems, the current density and conversion efficiency are calculated as functions of the bias voltage using the iterative method as well as the detailed balance approach (with or without a constant photon chemical potential) to demonstrate the significance of photon chemical potential on the performance of near-field TPV systems.

## II. METHODOLOGY

A typical near-field TPV system consists of an emitter and a cell with an electric circuit, as shown in Fig. 2 for a one-dimensional (1D) multilayer structure. There are $L + 1$ layers: $m = 0, 1, \ldots L$, and the end regions are assumed to be semi-infinite. A nanoscale vacuum gap ($d$) separates the emitter and the cell. While the method presented here is general, an InAs $p$-$n$ junction is chosen as the active region due to its narrow bandgap ($E_g = 0.354$ eV at room temperature) and high quantum efficiency [25, 26]. The p-doped and n-doped InAs layers with thicknesses $d_p$ and $d_n$ are subdivided by a nonuniform mesh to determine the local photogeneration rate and to solve the charge transport equations. Both sides of the p-n junction are partially or completely coated with metal films acting as electrodes. The backside Au coating thickness $d_{Au}$ is chosen to be 100 nm to improve the efficiency through reflection of sub-bandgap photons while serving as the back electrode [27]. The front metal grid is ignored in the model due to low shading area. An ITO thin film is coated on a semi-infinite bulk tungsten as a frequency-tunable Drude emitter, which can be optimized to match the bandgap of InAs to enhance the photogeneration rate [19]. The thickness of the ITO film is fixed to $d_{ITO}$ = 30 nm for optimized performance according to [6].

**A. Photon transport modeling**

A multilayer fluctuational electrodynamics formalism is applied to calculate photon exchange between any two layers in this 1D stratified medium. The net photogeneration rate per unit area in layer j of the cell ($3 \leq j \leq L-2$) is calculated by [8]

$$G_j = \sum_{m=0}^{L} \int_{\omega_g}^{\infty} \left[ \Psi(\omega, T_m, \mu_m) - \Psi(\omega, T_j, \mu_j) \right] \Upsilon_{mj}(\omega) d\omega \qquad (1)$$

In Eq. (1), $\omega$ is the angular frequency, $\omega_g = E_g/\hbar$ is the frequency that corresponds to the bandgap energy ($E_g$) with $\hbar$ being the reduced Planck constant, and $T$ and $\mu$ are the absolute temperature and photon chemical potential of the corresponding layer, respectively. The function $\Psi$ is the modified Bose-Einstein distribution, which is expressed as [3, 4]

$$\Psi(\omega,T,\mu) = \begin{cases} \left[\exp\left(\dfrac{\hbar\omega}{k_{\mathrm{B}}T}\right)-1\right]^{-1}, & \omega < \omega_{\mathrm{g}} \\ \left[\exp\left(\dfrac{\hbar\omega-\mu}{k_{\mathrm{B}}T}\right)-1\right]^{-1}, & \omega \geq \omega_{\mathrm{g}} \end{cases} \quad (2)$$

where $k_{\mathrm{B}}$ is the Boltzmann constant. The function $\Upsilon_{mj}$ is the fraction of photons emitted at a given frequency from layer $m$ that is absorbed by layer $j$ and vice versa. It can be calculated from the fluctuational electrodynamics as follows.

$$\Upsilon_{mj}(\omega) = \left(\frac{\omega}{c\pi}\right)^2 \mathrm{Re}\left\{i\,\mathrm{Im}(\varepsilon_m)\int_0^\infty k_\parallel dk_\parallel \right. \\ \left. \times \int_{z_{m-1}}^{z_m}\left[F(\omega,k_\parallel,z,z_{j-1})-F(\omega,k_\parallel,z,z_j)\right]dz\right\} \quad (3)$$

Here, $c$ is the speed of light in vacuum, $i=\sqrt{-1}$, $\varepsilon_m$ is the relative permittivity of layer $m$, and $k_\parallel = \sqrt{k_x^2 + k_y^2}$ is the parallel wavevector component. The function $F$ needs to be evaluated based on the Weyl components of the dyadic Green's electric and magnetic tensors. The formulation has been detailed in Refs. [18, 28] and will not be repeated here. In essence, it represents a photon transmission probability between two locations [8]. Note that $\Upsilon$ depends on the vacuum gap distance $d$, and is reciprocal, *i.e.*, $\Upsilon_{mj} = \Upsilon_{jm}$.

The rate of the net absorbed energy per unit area of the cell from the emitter (*i.e.*, net heat flux) is calculated from [8]

$$Q = \sum_{m=0}^{1}\sum_{j=3}^{L-2}\int_0^\infty \hbar\omega\left[\Psi(\omega,T_m,\mu_m)-\Psi(\omega,T_j,\mu_j)\right]\Upsilon_{mj}(\omega)d\omega \quad (4)$$

A recursive transfer approach is used in this work to solve the amplitude of electric and magnetic fields in each layer [29], which are then used to calculate the Weyl components [28].

The recursive approach, similar to the scattering matrix method, has the advantage of eliminating the exponential growing terms that may prevent the transfer matrix method from obtaining convergence [8, 28].

**B. Charge transport modeling**

Electrons and holes are the two types of charge carriers moving inside a semiconductor device. The conservation of charge number is prescribed by the continuity equations for electrons and holes expressed as [30, 31]

$$\frac{\partial n}{\partial t} = \frac{1}{q} \nabla \cdot J_e + g - r_b \tag{5}$$

$$\frac{\partial p}{\partial t} = -\frac{1}{q} \nabla \cdot J_h + g - r_b \tag{6}$$

where $n$ and $p$ are the electron and hole concentrations, which are functions of time and space, $q$ is the elementary charge, $g$ is the net generation rate that is obtained by $G_j / \Delta z_j$ in a given layer $j$, and $r_b$ is the (bulk) nonradiative recombination rate, which is the sum of the Auger recombination rate and the Shockley-Read-Hall (SRH) recombination rate, expressed respectively in the following [30, 32]:

$$r_{\text{Auger}} = (C_e n + C_h p)(np - n_i^2) \tag{7}$$

and

$$r_{\text{SRH}} = \frac{np - n_i^2}{\tau_p (n + n_{t,b}) + \tau_n (p + p_{t,b})} \tag{8}$$

In Eqs. (7) and (8), $C_e$ and $C_h$ are the Auger recombination coefficients for electrons and holes, $n_i$ is the intrinsic carrier concentration, $\tau_n$ and $\tau_p$ are the bulk lifetimes for electrons and holes, respectively, $n_{t,b}$ and $p_{t,b}$ are the electron and hole trap concentrations that are set to be the same as the $n_i$ in the modeling. The charge current densities, $J_e$ and $J_h$, are modeled in terms of drift and diffusion forces as follows

$$J_e = -q\upsilon_e n \nabla\varphi + qD_e \nabla n \tag{9}$$

$$J_h = -q\upsilon_h p \nabla\varphi + qD_h \nabla p \tag{10}$$

where $\upsilon_e$ and $\upsilon_h$ are the mobility of electrons and holes, respectively, and $D_e$ and $D_h$ are the diffusion coefficients, which are related to mobility according to Einstein's relation $D = \upsilon k_B T / q$ for each type of carrier. In Eqs. (9) and (10), $\varphi$ is the electrostatic potential, which obeys Poisson's equation:

$$\nabla \cdot (\varepsilon_r \nabla\varphi) = (q/\varepsilon_0)(N_A - N_D + n - p) \tag{11}$$

where $\varepsilon_r$ is the dielectric constant of the cell material, $\varepsilon_0$ is the vacuum permittivity, and $N_A$ and $N_D$ are the acceptor and donor concentrations, respectively. In the *p*-doped region, $N_D = 0$ and $N_A = $ const. In the *n*-doped region, $N_A = 0$ and $N_D = $ const.

Substituting Eqs. (1) and (7) through (10), into the three governing equations (5), (6) and (11) for prescribed conditions gives three differential equations in terms of the three unknown variables *n*, *p*, and $\varphi$. Obtaining stable and converging solutions is not a trivial task especially with high injection [24]. Here, a finite-difference method applying the Scharfetter-Gummel discretization scheme is used to solve the transient charge transport model with given boundary conditions and the initial guess obtained under the equilibrium condition [32-34]. In the present study, the spatial dependence is along the $z$ direction only and thus it is a 1D problem. This model considers both the majority and minority carrier concentrations, which provide detailed information beyond the depletion approximation typically used to determine the minority carrier concentrations in modeling PV and TPV systems [13, 16, 27, 35-37].

Boundary conditions for both sides of a *p-n* junction are required to solve the carrier transport equations. For a semiconductor device, the interface between semiconductor and metal is the most complicated part to model, which depends on material properties and operating conditions [38]. Surface passivation and selective contacts have often been used to boost the performance [39, 40]. Ohmic contact is assumed and simplified boundary conditions are used to describe the surface recombination [30, 38, 41]:

$$J_e(z_2) = qS_{e,p}\left[n(z_2) - n_0(z_2)\right] \tag{12}$$

$$J_h(z_2) = -qS_{h,p}\left[p(z_2) - p_0(z_2)\right] \tag{13}$$

$$J_e(z_{L-2}) = -qS_{e,n}\left[n(z_{L-2}) - n_0(z_{L-2})\right] \tag{14}$$

$$J_h(z_{L-2}) = qS_{h,n}\left[p(z_{L-2}) - p_0(z_{L-2})\right] \tag{15}$$

where $S_{e,p}$, $S_{h,p}$, $S_{e,n}$, and $S_{h,n}$ are the surface recombination velocities for electrons and holes in the *p* and *n* regions, respectively, and $n_0$ and $p_0$ are the carrier concentrations at equilibrium. Note that $n_0$ and $p_0$ are related to the intrinsic carrier concentrations and the acceptor concentrations (in the *p*-region) or the donor concentrations (in the *n*-region).

The obtained electron or hole concentration is related to electron or hole quasi-Fermi levels ($E_{f,e}$ or $E_{f,h}$) using the Boltzmann approximation:

$$n = n_i \exp\left(\frac{E_{f,e} - E_i}{k_B T}\right) \tag{16}$$

or
$$p = n_i \exp\left(\frac{E_i - E_{f,h}}{k_B T}\right) \tag{17}$$

where $E_i$ is the intrinsic Fermi energy. The quasi-Fermi levels obtained from Eqs. (16) and (17) are used to calculate photon chemical potential,

$$\mu(z) = E_{f,e}(z) - E_{f,h}(z), \quad z_2 \leq z \leq z_{L-2} \tag{18}$$

**C. Iterative solver of the coupled models**

Since the profile of photon chemical potential is an input in Eqs. (1) and (4), the photon and charge transport equations are coupled. A general iterative solution algorithm is outlined here. To begin with, an assumed profile of photon chemical potential is used to calculate the photogeneration profile in the cell according to fluctuational electrodynamics. Then, the quasi-Fermi level split is found by solving the semiconductor charge transport equations with the calculated photogeneration profile. This process yields an updated profile of the photon chemical potential. These steps are repeated until the specified convergence criteria are met. In the present modeling, it is assumed that the emitter and the cell are each at thermal equilibrium with specified temperatures $T_e$ and $T_c$. This iterative solution method was also introduced and validated by Callahan *et al.* [42] using an open-source solver [41] of the charge transport equations and extended further to the study of thermoradiative cells.

The general procedure of this iterative solver is described in the following.

1. <u>Initialization</u>. Specify the system parameters, such as geometry, vacuum gap, the emitter and cell materials and their electrical and optical properties. The intrinsic, donor, and acceptor concentrations and recombination coefficients are also needed for the charge transport calculations. A suitable mesh size must be chosen especially for the cell regions.

2. <u>Specification</u>. Specify a bias voltage, $V$ as the boundary condition. As a default, set $V = 0$ (short-circuit case) to begin.

3. <u>Trial input</u>. Set the photon chemical potential $\mu(z) = 0$ everywhere. This serves as the algorithm's initial input of the chemical potential profile.

4. <u>Photogeneration calculation</u>. Obtain the photogeneration profile based on Eq. (1) using the recursive transfer approach under the framework of fluctuational electrodynamics for specified chemical potential profile $\mu(z)$. Note that the obtained generation rate is assumed to be uniform within each layer $\Delta z_j$ $(j = 3,...L-2)$. This is exactly what is needed in the numerical solution of the charge transport equations.

5. <u>Charge transport solution</u>. Solve the steady-state charge transport model using the transient scheme based on the photogeneration profile obtained from the previous step. Doing so will yield values $E_{f,e}$ and $E_{f,h}$ and thus an updated $\mu(z)$ according to Eq. (18).

6. <u>Iteration.</u> Repeat steps 4 and 5 until the convergence criteria are met. In the present study, an upper limit in the relative difference between the photogeneration rate and an upper limit for the absolute difference between the photon chemical potential are used in combination to yield satisfactory convergence within reasonable computational time.

7. <u>Current and power outputs:</u> The current density in the circuit for the given bias voltage can be calculated using $J = J_e + J_h$ in the junction since the current density is the same around the circuit at steady state.

The voltage-dependent power per unit area and conversion efficiency are given by

$$P(V) = J(V)V \qquad (19)$$

and
$$\eta(V) = P(V)/Q(V) \qquad (20)$$

The net heat flux $Q$ depends on the photon chemical potential profile which, in turn, is a function of the bias voltage.

The above procedure yields the complete solution that can be used to evaluate the performance of the TPV system at a certain forward bias voltage. The current density and voltage characteristic curve (*i.e.*, the *J-V* curve) are produced by repeating steps 2-7 with appropriate increment in *V* from the short-circuit condition ($V = 0$) to the open-circuit condition when $J = 0$. Some trial-and-error are needed to determine the open-circuit voltage ($V_{oc}$) by refining the step size. It takes about 2-3 days to run a *J-V* curve for a given case with a dual eight core XEON E5-2687W 3.1GHz workstation using parallel computing. It should be noted that the inclusion of chemical potential in the photogeneration and radiative heat transfer based on fluctuating electrodyanmics has taken into account the luminescence and photon recycling effects [18].

**D. Detailed balance analysis**

For comparison to the iterative solution, the detailed balance analysis, originally developed by Shockley and Queisser [43] to estimate the efficiency limit of PV cells, is revisited by considering surface recombination. Here, the carrier concentrations are assumed to be uniform in each of the *p*-doped and the *n*-doped regions. Applying this procedure to near-field TPV cells, the current density is calculated from

$$J = q\left(G_{\text{cell}} - R_{\text{cell}}\right) \quad (21)$$

Here, $G_{\text{cell}}$ is the net photogeneration rate in the active region per unit area and is calculated by the summation:

$$G_{\text{cell}} = \sum_{j=3}^{L-2} G_j \quad (22)$$

In Eq. (21), $R_{\text{cell}}$ is the recombination rate per unit area that includes both bulk and surface recombination according to

$$R_{\text{cell}} = \left(r_{\text{Auger},p} + r_{\text{SRH},p}\right)d_p + R_{\text{s},p} + \left(r_{\text{Auger},n} + r_{\text{SRH},n}\right)d_n + R_{\text{s},n} \quad (23)$$

In the detailed balance analysis, it is assumed that $p = p_0 + \Delta$ and $n = n_0 + \Delta$, where $\Delta$ depends on the bias voltage and temperature according to [30, 43]:

$$pn = p_0 n_0 \exp(Vq/k_{\text{B}}T) \quad (24)$$

Equation (24) can be solved to obtain *p* and *n* either in the *p*-region or the *n*-region of the TPV cell, and then used to compute the bulk recombination rates from Eqs. (7) and (8).

The surface recombination rates per unit area are given as [30, 32]

$$R_{\text{s},p} = \frac{np - n_{\text{i}}^2}{(n + n_{\text{t,s}})/S_{\text{h},p} + (p + p_{\text{t,s}})/S_{\text{e},p}} \quad (25)$$

$$R_{s,n} = \frac{np - n_i^2}{(n + n_{t,s})/S_{h,n} + (p + p_{t,s})/S_{e,n}} \tag{26}$$

In the modeling, the surface trap concentrations are assumed to be the same as the intrinsic carrier concentration, *i.e.*, $n_{t,s} = p_{t,s} = n_i$. The surface recombination velocities are the same as those used in the iterative model.

Substituting Eq. (21) into Eq. (19) gives the power generation per unit area, which is a function of the bias voltage. Then Eq. (20) is used to calculate the efficiency in the detailed balance analysis. However, in applying Eq. (1) to calculate the net photogeneration rate and Eq. (4) for the net absorbed energy rate, one may set the photon chemical potential either as a constant ($\mu = Vq$) to approximate the luminescence effect or as zero ($\mu = 0$) to ignore the luminescence effect. In the detailed balance analysis, the photon transport and charge transport equations are decoupled and, therefore, they can be independently solved. In the following, these two scenarios ($\mu = Vq$ and $\mu = 0$) are used to compare with the solutions obtained from the iterative method in order to demonstrate the importance of considering the photon chemical potential as well as its spatial variation.

## III. RESULTS AND DISCUSSION

A simple analysis is given first to illustrate the effect of photon chemical potential without using the iterative solution or the detailed balance approach. Then, two InAs cells with different thicknesses are modeled with the iterative method and compared with the detailed balance analysis under either a constant or zero photon chemical potential assumption.

## A. Simple analysis of the luminescence effect in TPV cells

The photon chemical potential is the characteristic parameter for the luminescence effect in TPV cells operating in either the far field or the near field. This can be evaluated by comparing the difference in net radiation exchange between the emitter and cell with and without a constant chemical potential applied to the cell. The most important photons for semiconductor materials are those close to the bandgap energy. For a simple analysis, one may consider photons at the frequency $\omega_g$. A relative error $\chi_g$ in the net radiation exchange for photons at the bandgap energy is introduced to estimate the significance of photon chemical potential on the radiative energy exchange:

$$\chi_g(\omega_g, T_e, T_c, \mu_c) = \left| \frac{Q_g(T_e, T_c, \mu_c) - Q_g(T_e, T_c, 0)}{Q_g(T_e, T_c, \mu_c)} \right| \\ = \frac{\Psi(\omega_g, T_c, \mu_c) - \Psi(\omega_g, T_c, 0)}{\Psi(\omega_g, T_e, 0) - \Psi(\omega_g, T_c, \mu_c)} \quad (27)$$

where $T_e$ and $T_c$ are the temperatures of the emitter and cell, respectively, $Q_g$ is the spectral radiation exchange between the emitter and cell at $\omega_g$, and $\mu_c$ ($< E_g$) is the photon chemical potential of the cell and is taken as a constant. Note that Eq. (27) is applicable to both the far-field and the near-field regimes. For a cell with given temperature and bandgap, the relative error will increase if the emitter temperature $T_e$ decreases or if the photon chemical potential $\mu_c$ increases.

Taking InAs as the cell with $E_g = 0.354$ eV at $T_c = 300$ K, the effect of photon chemical potential is shown in Fig. 3a for various emitter temperatures. The error of radiation exchange between the emitter and cell due to neglecting the photon chemical potential exponentially increases as the photon chemical potential increases. On the other hand, if the emitter temperature is higher, a larger photon chemical potential may be tolerable with the same error bound. This can be seen clearly in Fig. 3b which plots the constant $\chi_g$ curves. For a given emitter temperature, the photon chemical potential must be small enough such that the error due to neglecting the photon chemical potential is below the error bound indicated by each curve. It should be noted that in a real TPV cell, the error in the total radiation exchange would be somewhat lower than $\chi_g$ since the effect of $\mu_c$ decreases as the frequency increases.

This simple analysis can be applied to semiconductors with different bandgaps. In Fig. 4, the 10% error curves are displayed for five common semiconductor materials (whose bandgap values are indicated in the figure) at $T_c = 300$ K. In order to maintain an error smaller than 10%, the photon chemical potential must not exceed certain values for a given $E_g$. Clearly, the photon chemical potential plays an important role, especially when the emitter is at moderate temperatures and the cell's bandgap is relatively small. While this simple analysis provides an intuitive suggestion about whether the iterative solution is necessary, it cannot quantify the actual luminescence effect. It will be shown in Sec. III.C that sometimes even though the photon chemical potential varies little throughout the *p-n* junction regions, its value may deviate from the product $Vq$.

**B. Parametrization of the near-field InAs systems**

Two near-field InAs TPV systems with different geometric structures are considered in this work. In the following, the emitter (layers $m = 0$ and 1) is assumed to be at a uniform temperature $T_e = 900$ K; the active region of the cell (layers $m$ from 3 to $L - 2$), the Au coating (layer $m = L - 1$), and air (layer $m = L$) are assumed to be at $T_c = 300$ K. The vacuum spacing is set to be $d = 10$ nm, which is sufficiently small to establish a high-injection level in the *n*-region. The dielectric function of tungsten is taken from the room temperature values [44] without considering temperature dependence. The dielectric function of InAs is obtained from Ref. [44] without considering the free-carrier contributions. Free-carrier contributions were included by Milovich *et al.* [26] using a Drude-Lorentz model, but this is not the main focus of the present study. The dielectric functions of ITO and Au are modeled by a Drude model $\varepsilon(\omega) = \varepsilon_\infty - \omega_p^2/(\omega^2 + i\gamma\omega)$ to account for free-electron contributions [8]. For ITO: $\varepsilon_\infty = 4$, $\omega_p = 1.52 \times 10^{15}$ rad/s, and $\gamma = 1.52 \times 10^{14}$ rad/s [6]; for Au: $\varepsilon_\infty = 1$, $\omega_p = 1.37 \times 10^{16}$ rad/s, and $\gamma = 5.31 \times 10^{13}$ rad/s [27]. The effects of temperature and lattice vibration on the dielectric function of ITO are neglected.

To model the InAs cell, the acceptor and donor concentrations are set to $N_A = 8 \times 10^{17}$ cm$^{-3}$ and $N_D = 2 \times 10^{16}$ cm$^{-3}$ in this study. The intrinsic concentration is $n_i = 6.06 \times 10^{14}$ cm$^{-3}$ for InAs at room temperature. For the bulk recombination, the following parameters are used: $C_e = C_h = 2.26 \times 10^{-27}$ cm$^6$ s$^{-1}$ and $\tau_e = \tau_h = 100$ ns [45].

A Caughey-Thomas-like model is used to describe the electron and hole mobilities depending on temperature and doping concentration [46]. This low-field mobility model is expressed as

$$\upsilon\left(N_{(A,D)}, T\right) = \upsilon_{\min} + \frac{\upsilon_{\max}(300/T)^{\theta_1} - \upsilon_{\min}}{1 + \left\{N_{(A,D)}/\left[N_{\text{ref}}(T/300)^{\theta_2}\right]\right\}^\zeta} \tag{28}$$

The parameter values for InAs are taken from [46]. For electrons, $\upsilon_{\min} = 1000$ cm$^2$V$^{-1}$s$^{-1}$, $\upsilon_{\max} = 34000$ cm$^2$V$^{-1}$s$^{-1}$, $N_{\text{ref}} = 1.1 \times 10^{18}$ cm$^{-3}$, $\theta_1 = 1.57$, $\theta_2 = 3.0$, and $\zeta = 0.32$; for holes, $\upsilon_{min} = 20 \text{ cm}^2\text{V}^{-1}\text{s}^{-1}$, $\upsilon_{max} = 530 \text{ cm}^2\text{V}^{-1}\text{s}^{-1}$, $N_{ref} = 1.1 \times 10^{17} \text{ cm}^{-3}$, $\theta_1 = 2.3$, $\theta_2 = 3.0$, and $\zeta = 0.46$.

In the following, a thin p-n junction and a thick p-n junction are taken as examples to demonstrate the nonuniform photon chemical potential and its effect on the performance of near-field TPV systems. For the thin cell, $d_p = d_n = 200$ nm, and the surface recombination velocities are set to be $S_{e,p} = S_{h,n} = 100$ cm/s and $S_{e,n} = S_{h,p} = 10,000$ cm/s to distinguish the minority carries and majority carriers. These values are within the range of high-quality films [26, 45]. For the thick cell, $d_p$ = 400 nm and $d_n$ = 5 μm, while all the surface recombination velocities are taken as $10^5$ cm/s, which is reasonable for typical semiconductor cells [45].

## C. Photon chemical potential profile and performance of a thin near-field InAs TPV system

For a conventional solar PV cell made of a p-n junction, the quasi-Fermi level split across the depletion region of the cell is assumed to be equal to the product of the forward bias voltage and elementary charge [26, 28, 29]. This is a widely recognized argument for PV cells with two hidden assumptions: the PV cells are under low-injection condition and the PV cells are thick. In most cases, the first assumption is likely to be valid for regular doped-silicon PV cells, because the injection level due to the spectral intensity from one or even several suns would still be moderate. For the second assumption, the widths of p- and n-regions in the PV cells are typically much larger than the depletion region. However, this may not be the case in near-field TPV systems with nanoscale separation distances, especially with very thin p-n junctions.

The band diagrams of the thin TPV cell ($d_p$ = $d_n$ = 200 nm) with a vacuum gap distance of $d$ = 10 nm are calculated at the short circuit and maximum efficiency (corresponding to $V$ = 0.17 V) conditions, as shown in Figs. 5a and 5b, respectively. The height of the shaded region represents the differences between the quasi-Fermi levels, which is nothing but the photon chemical potential profile $\mu(z)$. The depletion region is broadened and extended throughout the *n*-region, especially for the short circuit case. Surface recombination also plays an important role on the depletion region as can be seen from the inflection of the curve. Note that high-injection condition exists in the *n*-region. Fig. 5a shows that $\mu(z)$ is not zero but a strong function of *z*. High-quality surfaces with less surface defects for both hole and electron, as used here, trap the free carriers inside the *p-n* junction, giving rise to a non-zero quasi-Fermi level difference in the depletion region even with $V = 0$. When $V = 0.17$ V, as can be seen from Fig. 5b, $\mu(z)$ is approximately a constant which is 0.195 eV, which is about 15% greater than the product $Vq$.

The profile of photon chemical potential is calculated for different voltages and shown in Fig. 6. For small bias voltages, $\mu(z)$ decreases as *z* increases. When *V* increases to above 0.13 *V*, the variation of $\mu(z)$ becomes negligibly small. Nevertheless, the calculated $\mu(z)$ from the iterative method is always greater than $Vq$. The difference in the photon chemical potential calculated from the iterative solution and by assuming a constant value based on the detailed balance analysis is expected to impact the characteristics of the TPV cell.

The luminescence effect is demonstrated in Fig. 7a and 7b by plotting the net photogeneration rate and the net absorbed energy rate (per unit area) of the cell versus the bias voltage. As *V* increases, the luminescence effect reduces both $G_{cell}$ and $Q$, as compared to the case with $\mu = 0$, when luminescence effect is completely neglected. Since the iterative solution gives $\mu(z)$ greater than $Vq$, the curve with $\mu = \mu(z)$ is always the lowest. As shown in Fig. 3b, when $\mu < 0.15$ eV, $\chi_g$ is within 5% and the impacts of chemical potential on the photogeneration and heat transfer are relatively small. The differences become larger as *V* increases to above 0.12 V. The maximum relative error in $G_{cell}$ for $\mu = Vq$ is 21% taking the iterative method as the reference, which occurs at $V = 0.22$ V. At $V = 0.17$ V when the maximum efficiency occurs according to the iterative method, the relative error is 6.1% with $\mu = Vq$ and 9.7% with $\mu = 0$. Note that $\chi_g$ based on $\mu = 0.17$ eV is 6.3%, suggesting that $\chi_g$ gives a good indication when the profile of photon chemical potential should be considered. In terms of the net absorbed energy, the error for using $\mu = Vq$ is 3.1% and that for $\mu = 0$ is 4.9% at the maximum efficiency condition, $V = 0.17$ V. The smaller error in $Q$ is because sub-bandgap photons that contribute more than 50% of the heat transfer rate are not altered by the photon chemical potential, as suggested in Eq. (2).

The current density and efficiency are calculated as functions of the voltage as shown in Fig. 8 with the iterative method, $\mu = \mu(z)$, and compared with these obtained from the detailed balance analysis for $\mu = Vq$ and $\mu = 0$. The detailed balance approach overpredicts the current density in particular as $V$ increases to beyond 0.13 V. The difference between the current densities with $\mu = Vq$ and $\mu = 0$ is purely due to the net photogeneration rate as the recombination rates are the same according to the detailed balance analysis. Since the recombination rate increases with $V$, the current density drops to zero as $V$ increases to the open circuit voltage $V_{oc}$. The open circuit voltage for $\mu = Vq$ is slightly larger than that for $\mu = \mu(z)$ but much smaller than that for $\mu = 0$ as can be seen from Fig. 8a. As shown in Fig. 8b, the maximum efficiencies are 39.8% at $V = 0.205$ V for $\mu = 0$, 37.4% at $V = 0.19$ V for $\mu = Vq$, and 32.7% at $V = 0.17$ V for $\mu = \mu(z)$. These differences are due to the combination of effects caused by the profile of photon chemical potential and the underlying charge transport. The comparisons of the major parameters reveal that the accurate modeling of the photon chemical potential and the charge transport are critical to the prediction of the performance near-field TPV systems.

It should be noted that with even higher level of injection, the Boltzmann distribution used to approximate the Fermi-Dirac distribution of electron and hole systems may be questionable; hence, more comprehensive models considering degenerate semiconductors may be necessary to accurately describe the band diagrams. Such a situation does not occur for the cases studied in this work and therefore will not be further discussed.

**D. Photon chemical potential profile and performance of a thick near-field InAs TPV system**

A thick near-field InAs TPV cell with high surface recombination velocities is modeled to examine the spatial variation of $\mu(z)$ and the accuracy of the detailed balance analysis. The band diagrams of the thick cell with the same prescribed emitter and cell temperatures ($T_e$ = 900 K and $T_c$ = 300 K) and separation distance ($d$ = 10 nm) at short-circuit and open-circuit conditions are shown in Fig. 9a and 9b, respectively. Note that the width of the $p$-region is 400 nm and that of the $n$-region is 5 μm. A narrow depletion region is observed between the $n$- and $p$- regions, since the injection level is moderate in this case. Furthermore, the quasi-Fermi level split across the depletion region is close to $Vq$. However, the height of the shaded region varies greatly since photogeneration is highly nonuniform in the near-field regime [13, 37].

The spatial variation of the photon chemical potential is shown in Fig. 10 at different forward voltages. When the voltage is below 0.13 V, $\mu(z)$ decreases and reaches a minimum ($\approx Vq$) in the depletion region. In the $n$-region, $\mu(z)$ gradually increases to reach a maximum and then decreases toward the end. For the open-circuit case, $\mu(z)$ varies monotonically from 0.18 eV to 0.13 eV. Clearly, a constant $\mu$ assumption is not valid for near-field TPV systems even for moderate injection level.

The current density and efficiency are plotted in Fig. 11 against the voltage for the thick-cell TPV system. The open-circuit voltage is $V_{oc} = 0.145$ V with the detailed balance analysis and $V_{oc} = 0.179$ V with the iterative method. The difference between $\mu = Vq$ and 0 is less than 5% for $V < 0.14$ V and hence the two curves almost overlap with each other. In thiscase, the luminescence effect is negligibly small due to the small chemical potential. The detailed balance analysis underpredicts both $J$ and $\eta$ when $V > 0.075$ V. As shown in Fig. 11b, the maximum efficiency of 22.2% (at $V = 0.13$ V) calculated by the iterative method is much greater than 17.1% (at $V = 0.1$ V) obtained from detailed balance. The detailed balance analysis inherently assumes infinite charge diffusion coefficients and cannot capture the spatial variation of the charge concentration. Hence, the charge concentration at the surface is overpredicted, resulting in a higher surface recombination rate and a lower efficiency. It is important to consider the full charge transport processes in order to accurately predict the performance of near-field TPV systems.

For the thick cells under the short-circuit condition, $G_{cell} = 3.87 \times 10^{20}$ cm$^{-2}$ s$^{-1}$, which is 3% higher than that for the thin cell and $Q = 28.5$ W cm$^{-2}$, which is 5% more than that for the thin cell. These values are comparable between the thin and thick TPV systems. The lower $J$ and $\eta$ with increased bias for the thick TPV system is due to the higher recombination rate in the thicker InAs cell. The enhancement of performance using thin-film TPVs has been suggested by Tong et al. [47].

## IV. CONCLUSIONS

This work presents an iterative method to study near-field TPV systems that takes account of the luminescence effect based on fluctuational electrodynamics with a modified Bose-Einstein distribution along with the full drift-diffusion model of the charge transport in the *p-n* junction. The spatial variation of the photon chemical potential is illustrated with two near-field InAs TPV cells operating at a vacuum gap distance of 10 nm. The results show that the detailed balance approach by ignoring the luminescence effect ($\mu = 0$) or by using a constant photon chemical potential ($\mu = Vq$) cannot properly describe the characteristics of both the thin and thick TPV systems and could cause 10-30% error in efficiency and output power when compared to the iterative solutions. This study demonstrates that the photon chemical potential is an important parameter in near-field semiconductor radiative energy converters and need to be carefully considered when the emitter is at a moderate temperature. A simple criterion is also introduced based on the percentage error of the emitted photons at the bandgap frequency that may give an initial estimate of the influence of $\mu$ for different semiconductor materials, with various bandgaps, at prescribed emitter and cell temperatures. The accurate modeling of the spatial profile of photon chemical potential provides researchers with a better understanding of photon-charge interactions in semiconductor *p-n* junctions, and moreover, it will benefit the design and development of TPV systems operating at nanoscale separation distances.

**ACKNOWLEDGMENTS**

D.F. and Z.M.Z. would like to thank the support from the U.S. Department of Energy (DOE), Office of Science, Basic Energy Sciences (DE-SC0018369). This work was authored in part by the National Renewable Energy Laboratory (NREL), operated by Alliance for Sustainable Energy, LLC, for the DOE under Contract No. DE-AC36-08GO28308. E.J.T. was supported by the Laboratory Directed Research and Development (LDRD) Program at NREL. The views expressed in the article do not necessarily represent the views of the DOE or the U.S. Government. D.V. would like to acknowledge the financial support from NSF (ECCS 1542160).

**DATA AVAILABILITY**

The data that support the findings of this work are available upon request through the corresponding authors.

**Figures captions**

Fig. 1.  Band diagram of a TPV cell under illumination and bias, showing the conduction band ($E_c$), valence band ($E_v$), and quasi-Fermi levels whose difference equals the photon chemical potential.

Fig. 2.  Schematic of an InAs TPV cell with an ITO coated tungsten emitter.

Fig. 3.  (a) The error due to neglecting photon chemical potential in radiation exchange as a function of photon chemical potential for various emitter temperatures; (b) the contour plot of the relative error in a $\mu$-$T_e$ plane.

Fig. 4.  The 10% error criterion for various semiconductor materials whose bandgap values are shown in parentheses.

Fig. 5.  The energy band diagrams of the thin near-field InAs TPV cell under (a) short-circuit ($V = 0$ V) and (b) maximum-efficiency ($V = 0.17$ V) conditions.

Fig. 6.  The profile of photon chemical potential in the $p$-$n$ junction of the thin cell for various forward voltages.

Fig. 7.  (a) Net absorbed energy density and (b) net photogeneration density versus forward bias voltage under three different treatments of the photon chemical potential profile for a thin InAs TPV cell.

Fig. 8.  (a) Current density and voltage characteristics and (b) efficiency and voltage characteristics using three different treatments of the photon chemical potential profile for a thin InAs TPV cell.

Fig. 9.  The energy band diagrams of the thick InAs TPV cell under (a) short-circuit and (b) open-circuit ($V = V_{oc} = 0.179$ V) conditions.

Fig. 10.  The profile of photon chemical potential of the thick cell with various forward bias voltages.

Fig. 11.  (a) Current density and voltage characteristics and (b) efficiency and voltage characteristics using three different treatments of the photon chemical potential profile for a thick InAs TPV cell.

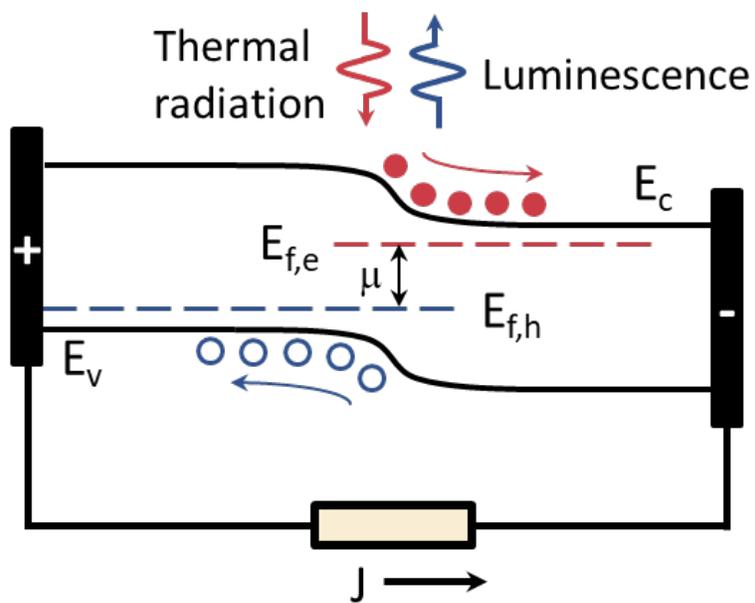

Fig. 1, Feng et al.

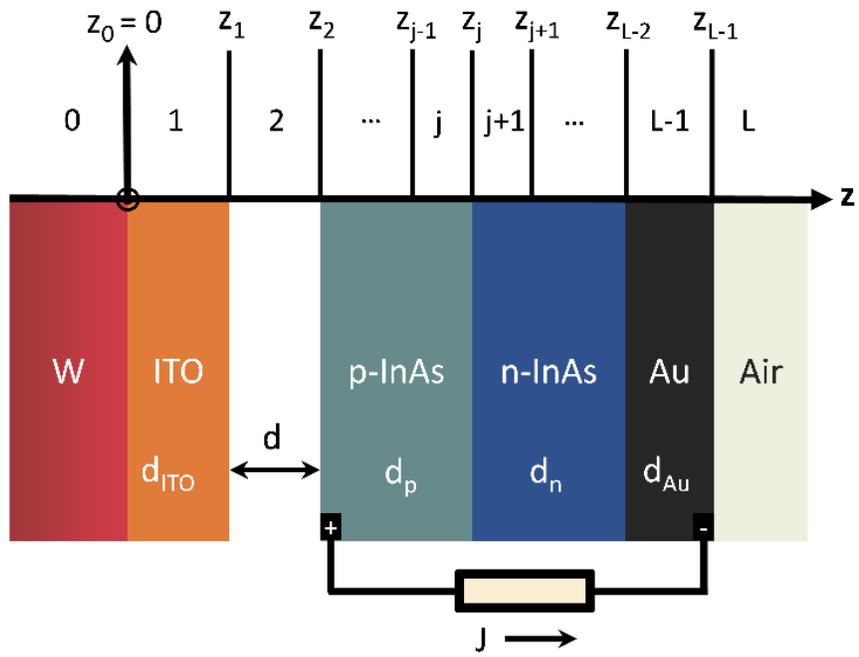

Fig. 2, Feng et al.

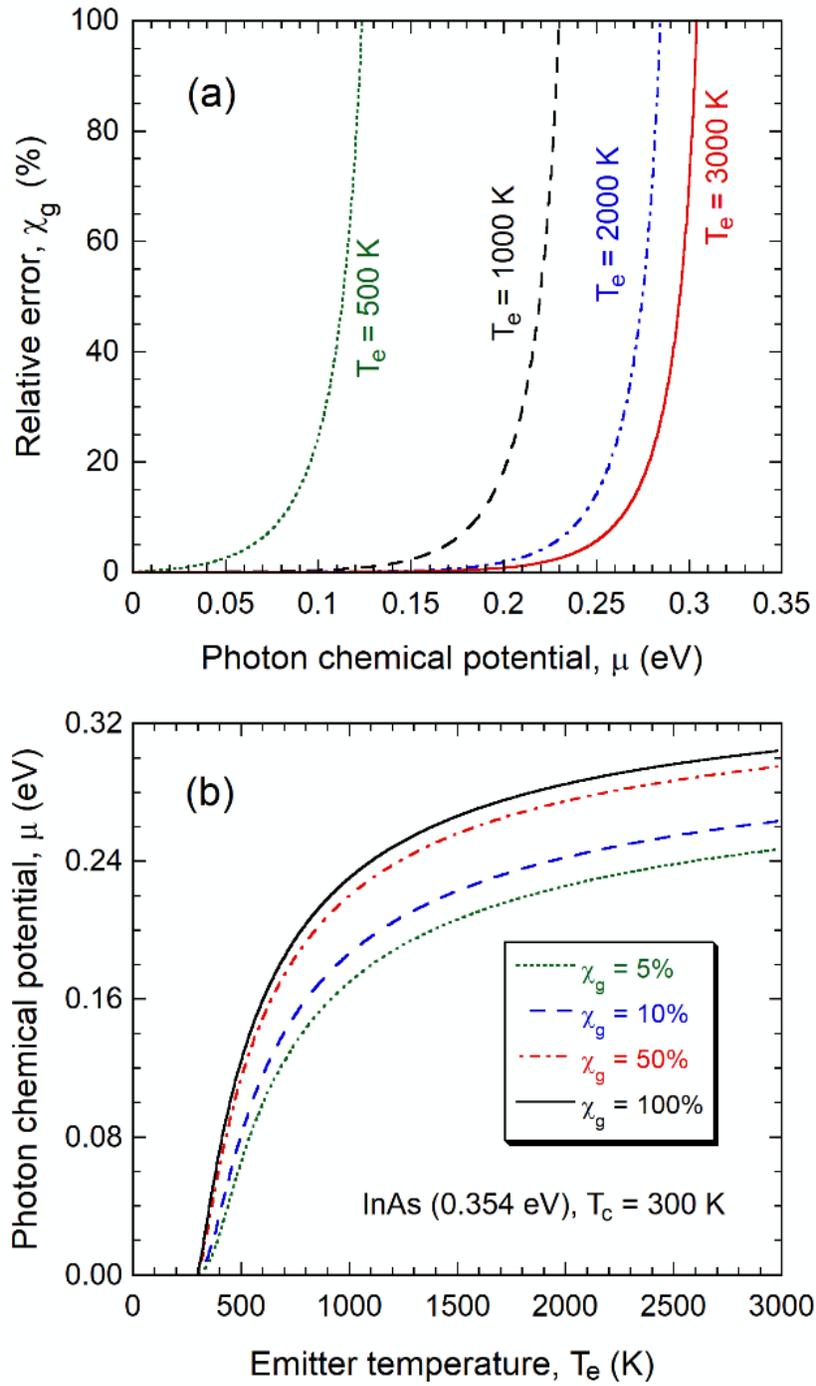

Fig. 3, Feng et al.

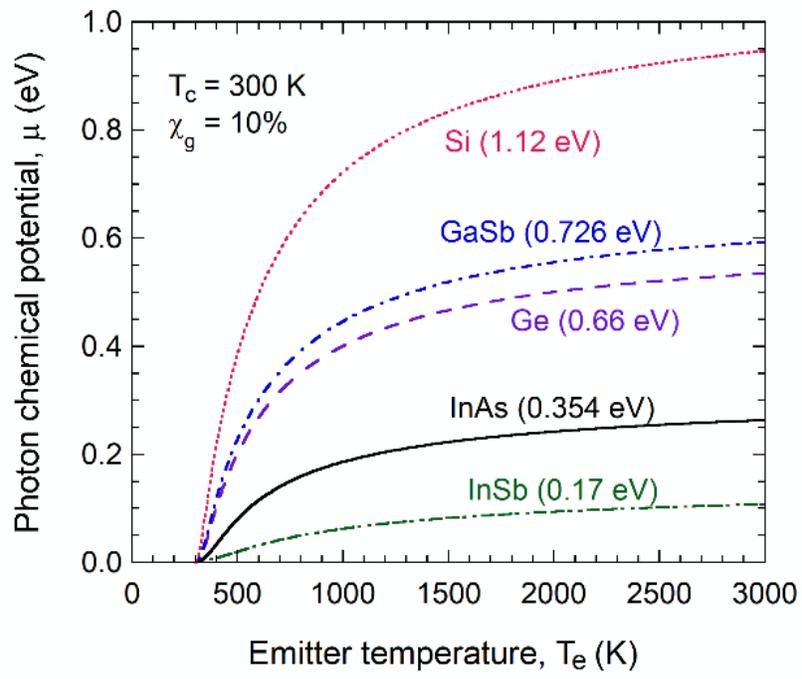

Fig. 4, Feng et al.

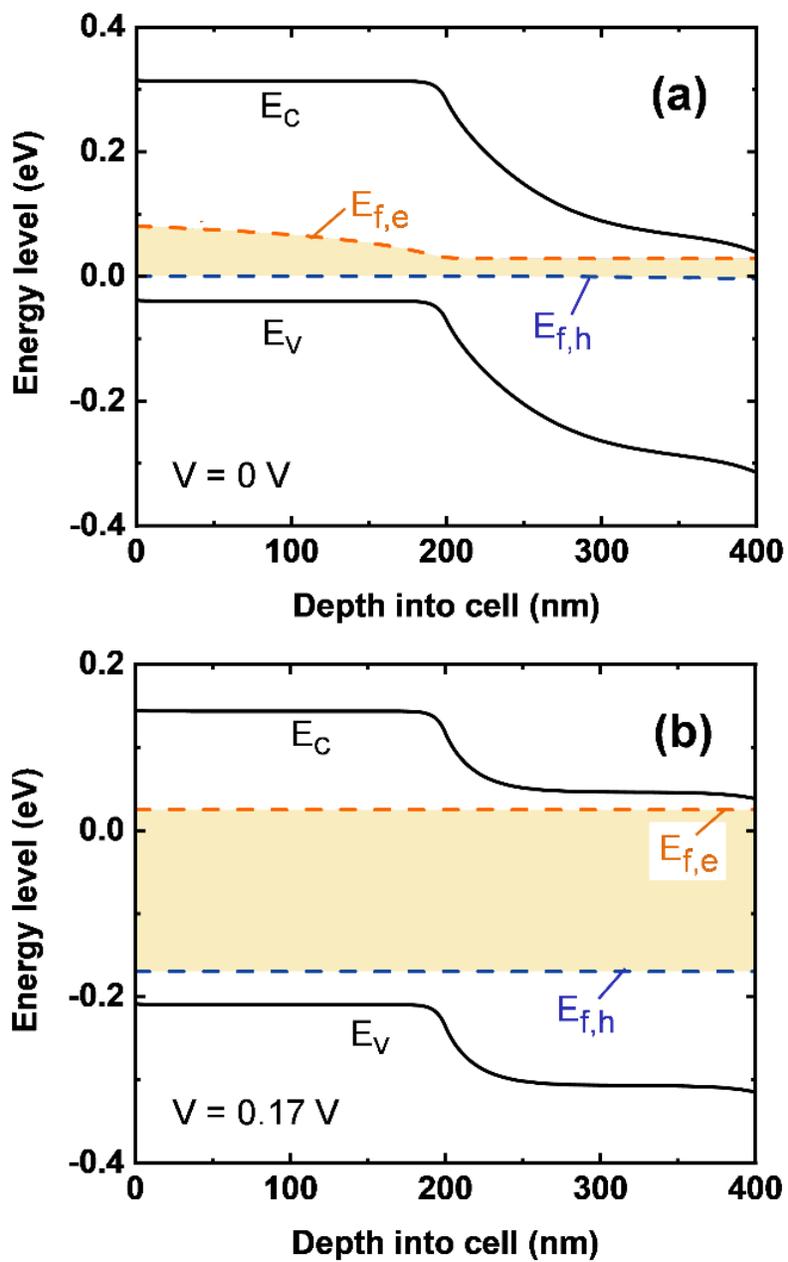

Fig. 5, Feng et al.

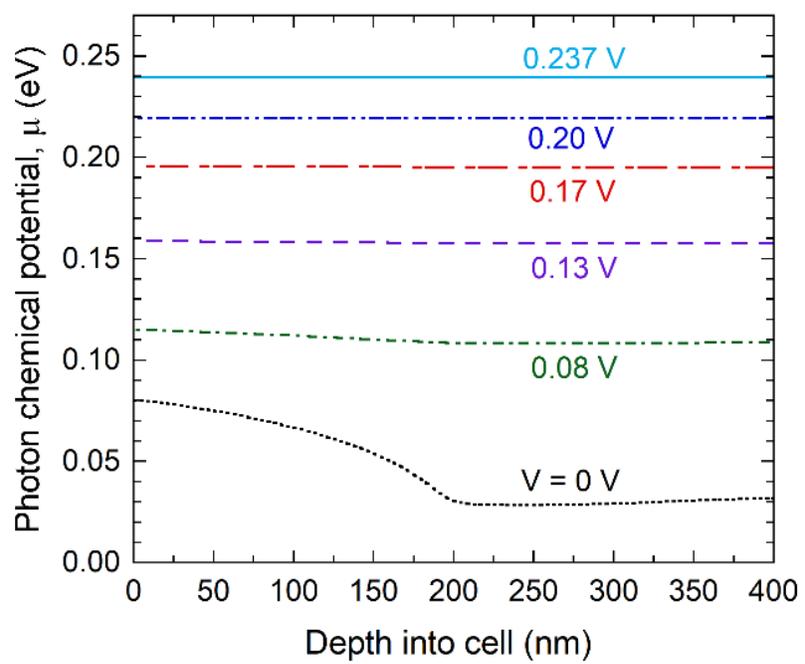

Fig. 6, Feng et al.

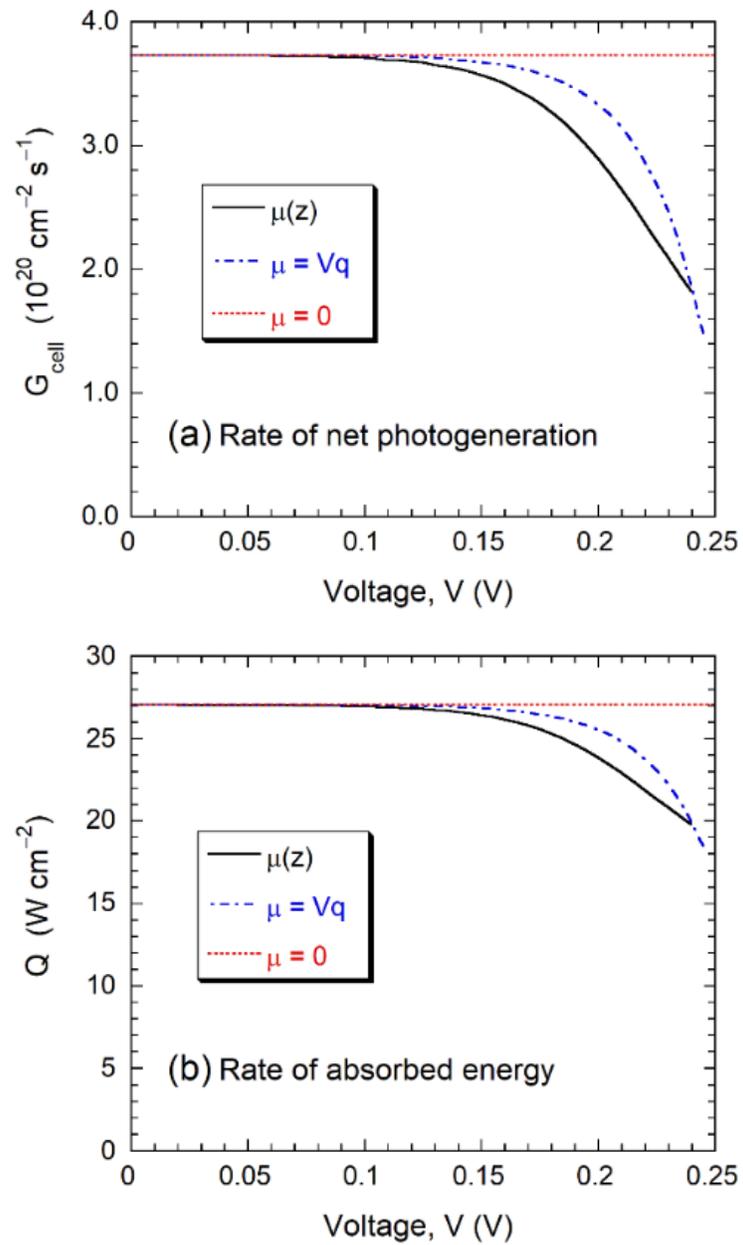

Fig. 7, Feng et al.

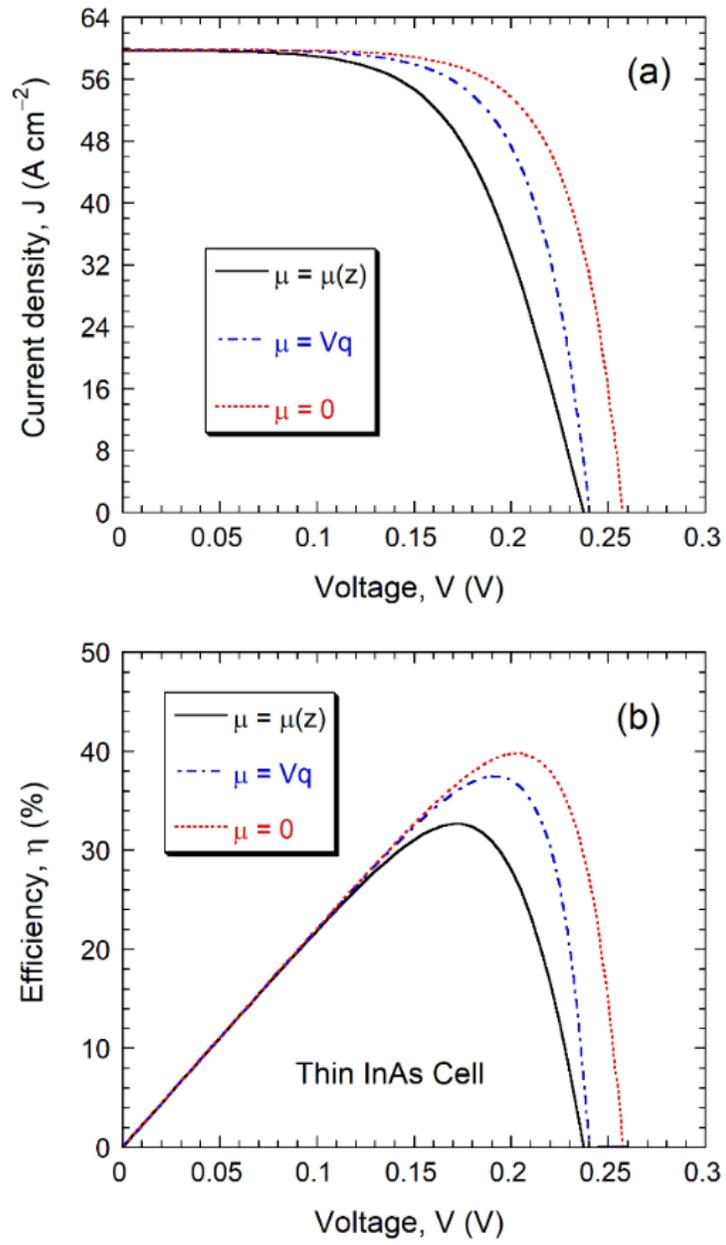

Fig. 8, Feng et al.

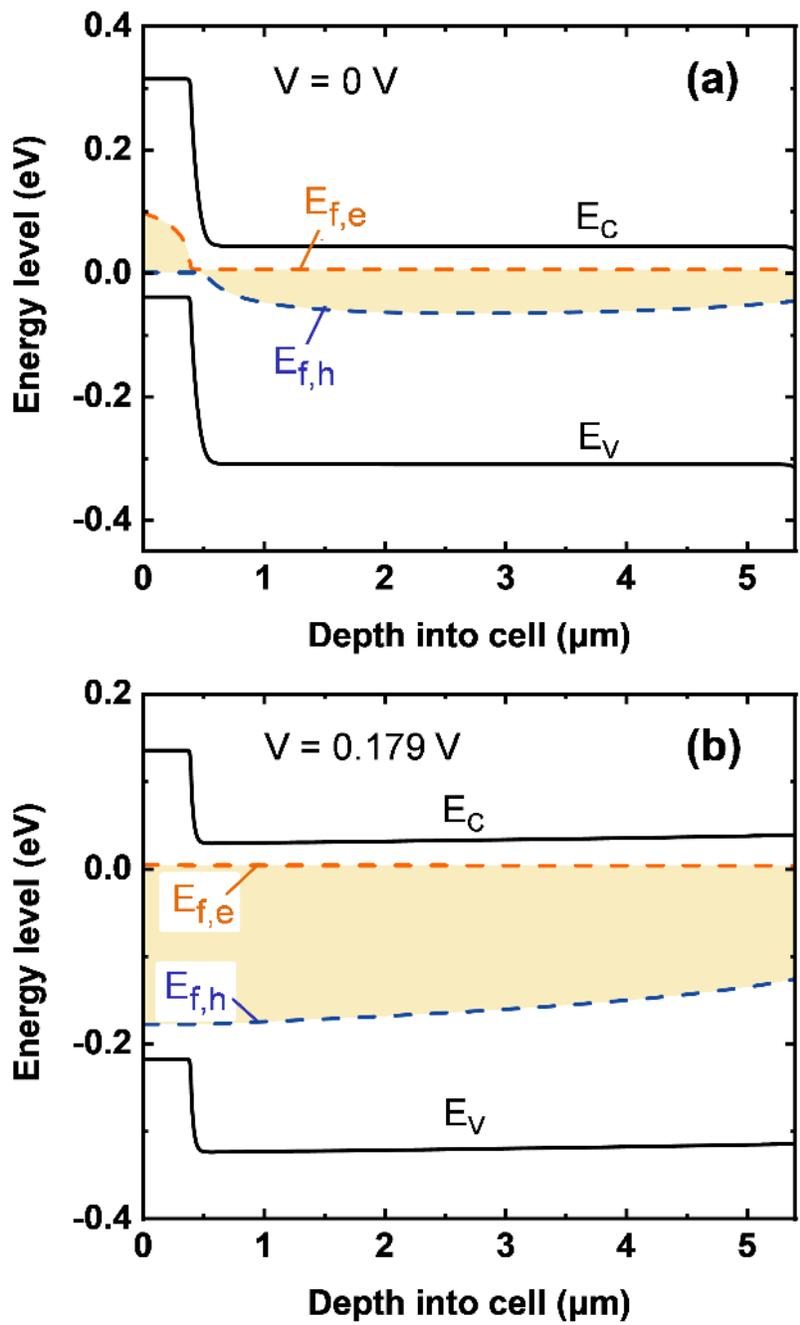

Fig. 9, Feng et al.

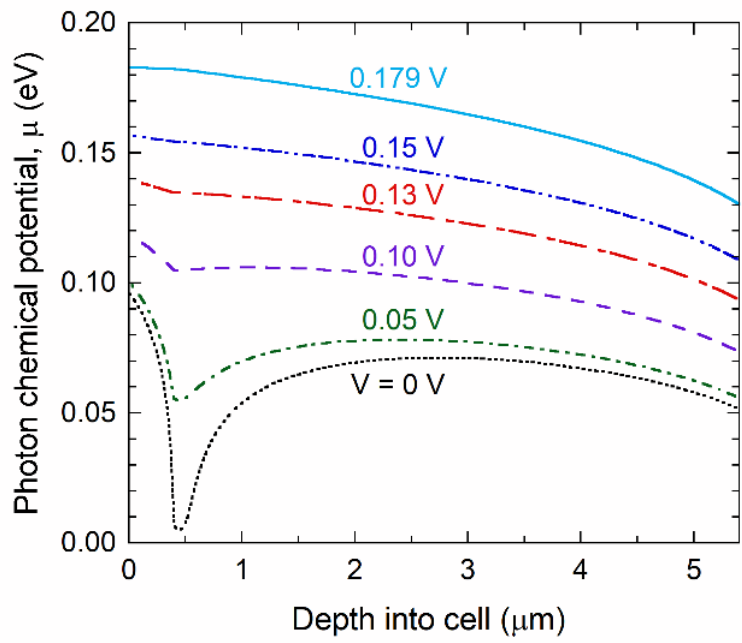

Fig. 10, Feng et al.

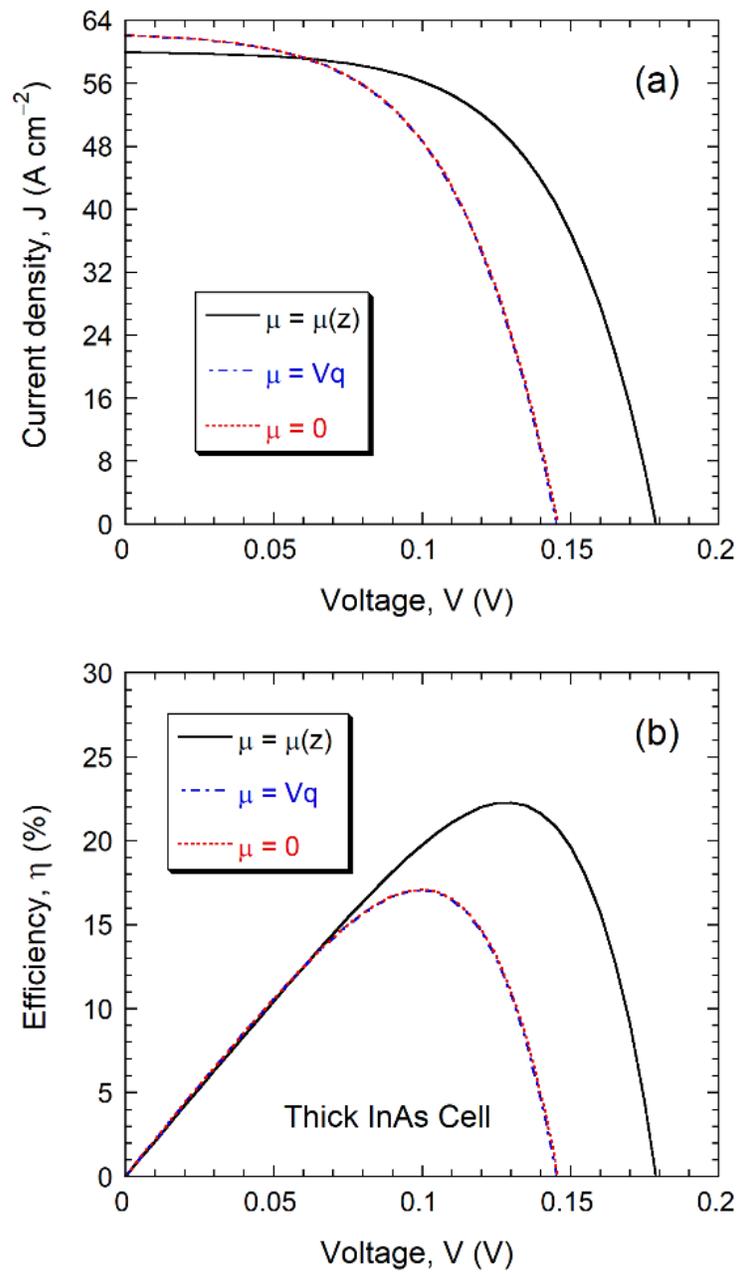

Fig. 11, Feng et al.